\documentclass[iop]{emulateapj}
\usepackage{apjfonts}
\usepackage{graphicx}

\newcommand{\be}{\begin{equation}}
\newcommand{\ee}{\end{equation}}
\newcommand{\bea}{\begin{eqnarray}}
\newcommand{\eea}{\end{eqnarray}}
\newcommand{\Msun}{M_{\odot}}
\newcommand{\lexp}{\mathop{\langle}}
\newcommand{\rexp}{\mathop{\rangle}}

\def\Msun{\ M_\Sol}

\def\rmFe{{\rm Fe}}
\def\kms{{\rm km\, s^{-1}}}

\def\facc{f_{\rm acc}}
\def\fagb{f_{\rm AGB}}
\def\esf{\epsilon_{\rm SF}}
\newcommand{\comment}[1]{}

\shortauthors{CONROY} 
\shorttitle{ON THE BIRTH MASSES OF THE ANCIENT GLOBULAR CLUSTERS}

\begin{document}
\journalinfo{The Astrophysical Journal}
\slugcomment{Accepted for Publication in ApJ}

\title{On The Birth Masses of the Ancient Globular Clusters}

\author{Charlie Conroy}
\affil{Harvard-Smithsonian Center for Astrophysics, Cambridge, MA,
  USA\\ Department of Astronomy \& Astrophysics,
  University of California, Santa Cruz, CA, USA}

\begin{abstract}

  All globular clusters (GCs) studied to date show evidence for
  internal (star-to-star) variation in their light element abundances
  (including Li, C, N, O, F, Na, Mg, Al, and probably He).  These
  variations have been interpreted as evidence for multiple star
  formation episodes within GCs, with secondary episodes fueled, at
  least in part, by the ejecta of asymptotic giant branch (AGB) stars
  from a first generation of stars.  A major puzzle emerging from this
  otherwise plausible scenario is that the fraction of stars
  associated with the second episode of star formation is observed to
  be much larger than expected for a standard IMF.  The present work
  investigates this tension by modeling the observed anti-correlation
  between [Na/Fe] and [O/Fe] for 20 Galactic GCs.  If the abundance
  pattern of the retained AGB ejecta does not depend on GC mass at
  fixed [Fe/H], then a strong correlation is found between the
  fraction of current GC stellar mass comprised of pure AGB ejecta,
  $f_p$, and GC mass.  This fraction varies from 0.20 at low masses
  ($10^{4.5}\Msun$) to 0.45 at high masses ($10^{6.5}\Msun$).  The
  fraction of mass associated with pure AGB ejecta is directly related
  to the total mass of the cluster at birth; the ratio between the
  initial and present mass in stars can therefore be derived.
  Assuming a star formation efficiency of 50\%, the observed Na-O
  anti-correlations imply that GCs were at least $10-20$ times more
  massive at birth, a conclusion that is in qualitative agreement with
  previous work.  These factors are lower limits because any mass-loss
  mechanism that removes first and second generation stars equally
  will leave $f_p$ unchanged.  The mass-dependence of $f_p$ probably
  arises because lower mass GCs are unable to retain all of the AGB
  ejecta from the first stellar generation.  Recent observations of
  elemental abundances in intermediate-age LMC clusters are
  re-interpreted and shown to be consistent with this basic scenario.
  The small scatter in $f_p$ at fixed GC mass argues strongly that the
  process responsible for the large mass loss is internal to GCs.  A
  satisfactory explanation of these trends is currently lacking.

\end{abstract}

\keywords{Galaxy: globular clusters --- globular clusters: general}


\section{Introduction}
\label{s:intro}

Evidence has been accumulating for the past thirty years that globular
clusters (GCs) harbor internal (star-to-star) variation in their light
element abundances \citep[including Li, C, N, O, F, Na, Mg, and
Al;][]{Cohen78, Kraft79, Smith82a, Smith83, Kraft94, Gratton04,
  Pasquini05, Smith05}.  And yet, save the most massive GCs, they can
still be considered mono-metallic in heavier elements including Ca,
Si, and Fe \citep[e.g.,][]{Carretta09a, Carretta10a}.

While early work focused on giant branch stars, abundance variations
have now been observed in main sequence turn-off stars
\citep[e.g.,][]{Gratton01, Briley02, Cohen02, Cannon98, Pancino10},
and so the observed variations cannot be attributed to non-canonical
mixing in evolved stars\footnote{Although non-canonical mixing cannot
  explain the totality of the observed star-to-star variation within
  GCs, there are well-documented correlations between some elemental
  abundances, such as carbon and nitrogen, and location along the
  giant branch \citep[see e.g.,][]{Smith02b} that cannot be explained
  by state-of-the-art stellar evolutionary models.  This fact
  significantly complicates interpretation of certain elements.}.
Rather, the observed star-to-star variation in elemental abundances
must be due to the fact that the stars formed from different material.

Perhaps the most striking result emerging from these observations is
that the number of stars within a GC that show anomalous abundance
ratios is comparable to those stars that have normal ratios.  In this
context `normal' abundance ratios refers to abundances characteristic
of field stars at the same [Fe/H] abundance, and `anomalous' refers to
abundance ratios that differ markedly from the field.  The comparable
number of normal and anomalous stars is observed for all GCs studied
to date, spanning a wide range in stellar mass and metallicity
\citep[e.g.,][]{Martell09, Carretta10c}.

The anomalous stars display abundance ratios that provide important
clues to the source of the raw material from which they formed.  These
stars show enhanced Na and Al and depleted O and Mg abundances.  They
are also CN-enhanced and CH-depleted.  These peculiar abundance
patterns arise naturally when matter is brought to very high
temperatures (far exceeding $10^7$ K).  At sufficiently high
temperatures the CNO, Na-Ne, and Mg-Al nuclear reaction cycles are
activated \citep[the precise temperature required for activation of
these cycles depends in detail on the site, whether e.g., the stellar
interior or the at the base of the convective
envelope;][]{Charbonnel06, Karakas07, Ventura08}.  In
intermediate-mass asymptotic giant branch (AGB) stars, the base of the
convective envelope reaches temperatures that are necessary to ignite
these nuclear cycles.  Because the envelope is convective, material
through the whole envelope can therefore participate in nuclear
burning.  This process is known as envelope, or hot bottom burning,
and it occurs for stars of initial masses in the range $4\Msun\lesssim
M\lesssim8\Msun$ \citep[e.g.,][]{Renzini81, Ventura08}.  Formation of
the anomalous abundance ratios within the envelopes of AGB stars is
appealing because these stars do not produce heavier elements such as
Ca, Si, and Fe, which then explains the lack of observed variation in
these elements in most GCs studied to date.  These stars also produce
large amounts of He, which seems necessary to explain the CMD
morphology of many GCs.

While AGB stars are plausible candidates for the source of the
peculiar abundance patterns, other sites have been proposed.
Alternatives include massive ($\gtrsim 20\Msun$) rotating stars
\citep{Decressin07} and massive binary stars \citep{deMink09}.  One of
the key differences between the massive star and AGB scenarios is the
timescales involved.  In the massive star scenario the peculiar
abundance patterns are created on a timescale comparable to the
production of type II supernovae (SNe).  This scenario therefore faces
two additional difficulties that the AGB scenario does not: 1) how to
retain the processed material within the natal GC in the face of
energy injection from SNe; and 2) how to create peculiar abundance
patterns that do not show evidence for type II SNe products such as Si
and Fe.  Other arguments against this scenario are discussed in
\citet{Conroy11b}.  While the AGB scenario is currently the favored
source of the peculiar abundance patterns, it should be noted that
this scenario contains significant uncertainties and theoretical
difficulties.

In light of these considerations, an intricate scenario has emerged to
explain the observations.  The key ingredient is that multiple
generations of star formation have occurred within GCs.  Later
generations of stars formed from gas enriched by the ejecta of AGB
stars associated with earlier generations of star formation.  Such a
scenario has been recently reviewed and discussed at length by
\citet{Conroy11b}, to which the reader is referred for details
\citep[see also, e.g.,][]{Cottrell81, Smith87, DErcole08, Renzini08,
  Carretta10c, DErcole10b}.  As discussed in that work, the number of
star formation events in typical GCs was probably limited to two --- a
first generation of stars that formed from material with abundance
patterns similar to field stars, and a second generation formed from
gas enriched by AGB ejecta from the first generation.  The timescale
between these two epochs of star formation is probably several $10^8$
yr, which is the timescale for several relevant physical processes
including AGB evolution, the time it takes for UV photon production to
drop enough to allow the gas to cool and form stars, and the onset of
prompt type Ia supernovae.  This small difference in age cannot be
seen in the main sequence turn-off point in the ancient GCs, but can
be observed in younger clusters.  Indeed, the massive intermediate-age
clusters in the Large Magellanic Cloud (LMC) have a spread in their
turn-off point that is consistent within an internal age spread of
several $10^8$ yr \citep{Goudfroij09, Milone09}, suggesting that this
scenario occurs at the present epoch as well as the distant past.

Another important ingredient in this scenario is that pure AGB ejecta
must be mixed with gaseous material that has abundance ratios similar
to the first generation \citep{Prantzos07, Ventura08b, Ventura09,
  DErcole10b}.  This is required to reproduce the observed range in
light element abundances, which extends smoothly from the very
anomalous to the normal.  A natural way to acquire additional material
with normal abundance ratios is via accretion from the ambient
interstellar medium \citep[ISM;][]{Pflamm-Altenburg09, DErcole10b,
  Conroy11b}, although other scenarios are possible \citep{DErcole08,
  Gratton10a}.  As demonstrated in \citet{Conroy11b}, significant
accretion from the ambient ISM is possible for the physical conditions
characteristic of young GCs (i.e., cold dense interstellar media and
low relative velocities).  Ram pressure is not important except for
the lowest masses, where indeed anomalous abundance ratios are not
observed, both in the Galactic open clusters \citep{deSilva09,
  Martell09, Pancino10b} and the intermediate-age clusters in the LMC
\citep{Conroy11b}.  Implicit in this requirement is that the accreted
material remain incompletely mixed with the AGB ejecta --- the stars
must form from material with a {\it range} of abundances, not just
some average of accreted material and AGB ejecta.

The greatest challenge facing this otherwise plausible scenario is in
explaining the roughly comparable number of first and second
generation stars, which is observed for GCs spanning a wide range in
mass.  Under the assumption that second generation stars form from AGB
ejecta plus a modest amount of accreted material, the ratio of first
to second generation stars should be of order 10:1.  That is, the
standard scenario predicts a number of second generation stars lower
by a factor of ten compared to observations.  The standard prediction
assumes 100\% star formation efficiency and is based on stellar
evolution theory and a canonical initial mass function (IMF), which
implies that only $\sim10$\% of the mass of a stellar population ends
up in AGB ejecta.  A variety of solutions to this problem have been
proposed, including a different IMF between first and second
generation stars \citep{Smith82b, Prantzos06}, and a substantially
larger mass at birth of all GCs \citep[][]{DAntona04, Bekki06,
  DAntona07, Decressin07b, DErcole08, Schaerer11}. The second scenario
requires that GCs were factors of $10-100$ more massive at birth.  If
this is correct, it constitutes a dramatic revision of our
understanding of GC formation and evolution.  This tension provides
the motivation for the present analysis.

In the present work it is assumed that AGB stars with masses in the
range $3-8\Msun$ contribute to the formation of the second generation
of stars.  However, the current generation of AGB nucleosynthetic
yields favors a narrower range of stellar masses, perhaps only
$5-8\Msun$, that are contributing to form the observed range of
abundance patterns of second generation stars
\citep[e.g.,][]{Ventura09, DErcole10b}.  Appealing to a smaller range
of masses only exacerbates the problem noted above because an even
larger initial GC is required to produce the same amount of polluted
material (a factor of two larger if one considers only $5-8\Msun$ AGB
stars, rather than $3-8\Msun$ AGB stars, everything else being equal).
Moreover, if only massive AGB stars are allowed to contribute, then
the timescale for the formation of second generation stars shrinks to
$<10^8$ yr.

If a second generation of stars forms within the potential well of a
first generation GC, one might expect the second generation stars to
be more spatially concentrated than the first generation, at least
initially.  $N-$body simulations have shown that if they begin more
centrally concentrated, second generation stars will remain more
concentrated than the first generation for a few central relaxation
times \citep{Decressin08, DErcole08}, after which time the relative
number of first and second generation stars becomes constant within
the half-mass radius.  There is evidence that the second generation is
more centrally concentrated than the first in NGC 1851
\citep{Zoccali09}, $\omega$Cen \citep{Pancino03, Sollima07}, and NGC
3201 \citep{Carretta10e}.  The situation for other GCs is less clear
\citep{Carretta09b}, owing to the small numbers of stars observed,
incomplete spatial coverage, and the short relaxation times for many
GCs.

Parallel to the advances in the chemical abundances of normal GCs has
been the revelation of distinct stellar populations in the
color-magnitude diagrams (CMDs) of the most massive GCs as revealed by
the {\it Hubble Space Telescope} \citep[see][for a review]{Piotto09}.
It is now clear that the massive GCs $\omega$Cen and NGC 2808 harbor
multiple distinct main sequences as seen in their CMDs.  Many more GCs
harbor distinct sub-giant branches and significant width in their red
giant branches.  Although it is clear that the spread in the CMDs is
related to the elemental abundance variations \citep{Yong08, Marino08,
  Carretta09b, Milone10}, a comprehensive understanding of the
relation between these two phenomena is currently lacking.  In
particular, while much attention has been focused on the peculiar CMDs
of the most massive GCs, it is far from clear that the features of
these GCs are characteristic of the population as a whole.

The present work aims to provide a quantitative understanding of the
relative frequencies of first and second generation stars, the amount
of material accreted from the ambient ISM, and the amount of AGB
ejecta required to explain the observed elemental abundance
variations, over a factor of 100 in GC mass.  This work therefore aims
to bridge our understanding of the properties of lower mass clusters
with the most peculiar massive GCs.  One of the major goals of the
present analysis is to answer the following question: ``how much more
massive did the progenitors of present day GCs need to be in order to
explain the observed abundance variations within GCs?''

\section{A Simple Model For the Abundance Variations}
\label{s:model}

One of the most intriguing observational results is the strong
anti-correlation between [Na/Fe] and [O/Fe] within a given cluster.
It has been shown in previous work that this anti-correlation can be
reproduced with a simple dilution model wherein second generation
stars form from normal and processed material mixed in varying amounts
\citep{Prantzos07, Ventura08b, Ventura09, DErcole10b}.  Processed
material means in this context matter that has been exposed to
temperatures exceeding $\approx7\times10^7K$, where the CNO and Na-Ne
nuclear reactions are active in the convective envelopes of massive
AGB stars.  Stars formed from pure processed material have enhanced
[Na/Fe] and depleted [O/Fe] abundances.  Normal material means here
matter that has the same abundance patterns as the initial stellar
population, and also probably has the same abundance patterns as field
stars in the Galaxy, at the same [Fe/H].  Normal material at the low
metallicities considered has enhanced [O/Fe] and either solar or
sub-solar [Na/Fe] abundances.  In the present discussion, `AGB ejecta'
will be synonymous with `pure processed material', as it is assumed
herein that the source of the processed material is AGB ejecta.

If the normal and processed abundances are denoted with subscripts
`o' and `p' and if the fraction of pure processed material in the
$j-$th star is $f_p^j$ then the abundance of element $i$ in a star
formed from a mixture of normal and processed material is simply:
\noindent
\be
[i/\rmFe]_j = {\rm log}\bigg( (1-f_p^j)10^{[i/\rmFe]_o} + 
f_p^j 10^{[i/\rmFe]_p} \bigg),
\ee
\noindent
assuming Fe is constant. 

It is important to notice that in the context of this model there will
be stars with $f_p^j=0.0$; these are truly first generation stars that
formed purely from normal material.  Subsequent generations of stars
then form from a mixture of AGB ejecta and normal material.  The
source of the normal material required to produce the observed
[Na/Fe] and [O/Fe] values in the second generation is currently
unknown.  A plausible source is accretion from the ambient ISM over
the several $10^8$ yr when intermediate-mass AGB stars are evolving,
and after the type II SNe have exploded and cleared the GC of any
remaining initial gas \citep{Conroy11b}.

Once the normal and processed abundances of [Na/Fe] and [O/Fe] are
specified, one can readily estimate $f_p^j$ for each star in a GC by
comparing the star's observed Na and O abundances to the model
predictions.  An estimate of the global value of $f_p$ can then be
obtained by averaging the individual $f_p^j$ values over all the stars
in the GC.  This global $f_p$ represents the fraction of the present
stellar mass comprised of pure AGB ejecta.  The fraction of
the present GC stellar mass formed from normal material is then
given by $1-f_p$.

This procedure for determining $f_p$ is straightforward.  An advantage
of this approach is that no arbitrary, sharp distinction is made
between first and later generations of stellar generations based on
their location in the Na-O plane.  This is advantageous because the
observed distribution of stars along the Na-O anti-correlation is
continuous in most GCs, and so there is no natural way to separate
first and later generations.

A complication in interpreting the results from such a dilution model
is that the fraction of normal material is not easily separable into
the normal material comprising the initial stellar population and the
normal material locked in second generation stars.  This ambiguity
arises because stars with $0<f_p\lesssim0.1$, which are second
generation stars, are confused with stars with $f_p=0.0$, which are
truly first generation stars.  The fraction of normal material that is
associated with second generation stars will be denoted by $\facc$.
The fraction of total mass in the first stellar generation is then
$f_1=(1-f_p)(1-\facc)$, and the fraction of total mass in accreted
material is $f_a=(1-f_p)\facc$.  The fraction of mass in the second
generation is then $f_2=f_a+f_p$.

Ultimately we are interested in knowing if the amount of mass in pure
AGB ejecta incorporated in second generation stars and given by $Mf_p$
where $M$ is the total GC stellar mass, can be produced from the first
stellar generation.  The initial mass needed to produce the observed
amount of AGB ejecta can be estimated as follows.  First, define
$\fagb$ as the fraction of initial mass that ends up in AGB ejecta and
is available for second generation star formation.  It can be computed
from stellar evolution in conjunction with an IMF.  Next, define
$\esf$ as the star formation efficiency, or the fraction of gas mass
within the young GC that ends up in second generation stars.  The
initial mass then needed to produce the observed amount of AGB ejecta
is $Mf_p/\fagb/\esf$.  The ratio between this quantity and the actual
amount of mass in long-lived\footnote{For a canonical \citet{Kroupa01}
  IMF, approximately 50\% of the initial stellar mass of a population
  is lost due to stellar evolution and death.  In the present analysis
  attention is focused on {\it long-lived stars}, so the massive stars
  that carry away substantial mass when they die are not included in
  bookkeeping except where explicitly mentioned.} first generation
stars, $M(1-f_p)(1-\facc)$, is:
\noindent
\be
\label{e:fm}
f_{M1} = max\bigg(1.0,\frac{f_p}{\esf\fagb(1-f_p)(1-\facc)}\bigg).
\ee
\noindent

The quantity, $f_{M1}$ is the mass enhancement factor for the first
generation, considering only long-lived stars.  If stars of all masses
are included then $f_{M1}$ should be increased by a factor of
$\approx2$ (see footnote 2).  The definition of $f_{M1}$ in Equation
\ref{e:fm} preserves the requirement that $f_{M1}$ cannot be less than
unity.  It is instructive to consider a quantitative example.  If
$\fagb\approx0.1$ (as suggested by stellar evolution and a canonical
\citet{Kroupa01} IMF), $f_p=0.5$, $\facc=0.5$, and $\esf=0.5$, then
$f_{M1}=40.0$.  This means that the first generation contained
$40\times$ more long-lived stars at birth compared to its present
mass.

In what follows a star formation efficiency of $\esf=0.5$ will be
adopted.  The simplest argument for a high star formation efficiency
in GCs is based on the Virial Theorem and the requirement that GCs
remain gravitationally bound.  This argument suggests that the total
mass in stars formed needs to be at least 50\% of the initial gas mass
\citep[the resulting bound mass depends on how quickly gas is removed
from the system, for details see][]{Lada84}.  The gas available for
second generation star formation finds itself in a unique
configuration in that it is at the bottom of a deep potential well
(set by the first generation stars).  In principle the star formation
efficiency could therefore be lower and yet still result in a bound
cluster.  The uncertainty carried by the parameter $\esf$ is an
inevitability of the present analysis.

While recognizing that in reality star formation is not 100\%
efficient, previous work has nonetheless made the assumption that
100\% of AGB ejecta from the first generation is incorporated into
second generation stars.  This assumption was made for the simple
reason that it minimizes the required mass enhancement factor.  In the
present work a more realistic value for $\esf$ has been adopted.

Another quantity of interest is the {\it total} mass enhancement
factor in long-lived stars, that is, the ratio of total mass at birth
to total present mass.  This quantity is
\noindent
\be
\label{e:ft}
f_t = max\bigg(1.0,\frac{f_p}{\esf\fagb}\bigg).
\ee
\noindent
Notice that this is smaller than $f_{M1}$ by a factor of
$(1-f_p)(1-\facc)$ for $f_t>1$.  In the example above, the total mass
enhancement factor would be 10, a factor of four smaller than the
mass enhancement factor for the first generation.  The former is
smaller than the latter because of the addition of second generation
stars, which at least partially compensates for the substantial loss
of first generation stars.  For reference, a GC evolving in isolation
that experiences no secondary generations of star formation and also
experiences no loss of stars from e.g., stellar evaporation or
ejection, will have $f_{M1}=1$ and $f_t=1$.

\begin{deluxetable}{lll}
\tablecaption{Summary of Symbols and Quantities} \tablehead{
  \colhead{} & \colhead{Derived From} & \colhead{Description}}
\startdata
$f_p$ & Na-O dilution model & fraction of GC mass comprised of 
AGB ejecta \\
$\facc$ & free parameter & fraction of normal material accreted \\
 & & \,\,\,from ambient ISM \\
$\fagb$ & stellar evolution, IMF & fraction of mass in AGB ejecta \\
 & & \,\,\,from a coeval stellar population \\
$\esf$ & free parameter &  efficiency of star formation \\
$f_a$ & $(1-f_p)\facc$ & fraction of GC mass in accreted material \\
$f_1$ & $(1-f_p)(1-\facc)$ & fraction of GC mass in first generation
stars \\
$f_2$ & $f_a+f_p = 1-f_1$ & fraction of GC mass in second generation
stars \\
$f_{M1}$ & Equation \ref{e:fm}  & mass enhancement factor for 
first generation \\
$f_t$ & Equation \ref{e:ft} & total mass enhancement factor
\vspace{0.1cm}
\enddata
\label{t:frac}
\end{deluxetable} 

Most authors assume that second generation stars are born with masses
in the range $0.1-0.8\Msun$ \citep[e.g.,][]{DErcole08, Lind11}.  This
assumption is made in order to minimize the required mass enhancement
factors.  If AGB ejecta is only placed in long-lived second generation
stars, then none of the ejecta is `wasted' on higher mass stars that
evolve and die on short timescales.  This assumption reduces the
required mass enhancement by a factor of $\approx2$ compared to the
assumption, made in the present work, that second generation stars
fully populate a standard \citet{Kroupa01} IMF.  A fully populated IMF
is assumed here because there is no physical reason to believe that
the IMF should be biased toward low masses during second generation
star formation.

An argument often made for an IMF of second generation stars that is
skewed toward low masses is that massive second generation stars would
explode and thereby truncate further star formation.  This is
certainly an issue that must be addressed in any comprehensive theory,
but it is not a strong argument for a skewed IMF.  As discussed in
\citet{Conroy11b}, the heating rate of UV photons should be
substantial during the first $\sim10^8$ yr of the young GC.  This
heating may be sufficient to delay star formation until the UV photon
production rate drops precipitously at $\sim10^8$ yr.  After this
time, the gas may cool catastrophically in a manner analogous to the
formation of the first generation of GC stars.  This simple scenario
suggests that massive second generation stars could form and yet not
have a debilitating effect on the conversion of a significant fraction
of the accumulated gas into stars.

The symbols and quantities defined in this section are summarized in
Table \ref{t:frac} for convenience.

\section{Constraints on the Birth Masses of Galactic GCs}
\label{s:res}

\begin{deluxetable*}{lclrrrlllllll}
\tablecaption{Summary of GC Data} \tablehead{
  \colhead{ID} & \colhead{} & \colhead{[Fe/H]} & 
\colhead{N} & \colhead{log(M)} & \colhead{[Na/Fe]$_{\rm o}$} &
\colhead{[O/Fe]$_{\rm o}$} & \colhead{[Na/Fe]$_p$}
& \colhead{[O/Fe]$_p$} &\colhead{$f_p$} & \colhead{$\lexp R\rexp/R_h$} &
\colhead{$f_{M1}$} & \colhead{$f_t$}\\ 
\colhead{} & \colhead{} & \colhead{(1)} & \colhead{(2)} &
\colhead{(3)} & \colhead{(4)} & \colhead{(5)} & \colhead{(6)} &
\colhead{(7)} & \colhead{(8)} & \colhead{(9)} & \colhead{(10)} & \colhead{(11)}}
\startdata
NGC 7099 &   M30 & -2.33 & 19 & 5.19 &-0.30 & 0.50 & 0.80 &-0.80 & 0.32 & 2.33 & 18.7 &6.37 \\
NGC 7078 &   M15 & -2.33 & 20 & 5.89 &-0.30 & 0.50 & 0.80 &-0.80 & 0.36 & 1.47 & 22.4 &7.19 \\
NGC 4590 &   M68 & -2.23 & 36 & 5.16 &-0.35 & 0.60 & 0.80 &-0.80 & 0.28 & 1.83 & 15.2 &5.52 \\
NGC 6397 &       & -1.98 & 13 & 4.87 &-0.30 & 0.40 & 0.80 &-0.80 & 0.20 & 1.04 & 10.1 &4.04 \\
NGC 6809 &   M55 & -1.98 & 75 & 5.24 &-0.25 & 0.45 & 0.80 &-0.80 & 0.33 & 1.14 & 19.9 &6.65 \\
NGC 6715 &   M54 & -1.57 & 76 & 6.23 &-0.15 & 0.40 & 0.80 &-0.80 & 0.42 & 3.65 & 29.5 &8.49 \\
NGC 1904 &   M79 & -1.55 & 39 & 5.37 &-0.25 & 0.30 & 0.80 &-0.80 & 0.31 & 1.92 & 17.9 &6.18 \\
NGC 6752 &       & -1.56 & 88 & 5.31 &-0.10 & 0.50 & 0.80 &-0.80 & 0.36 & 1.69 & 22.1 &7.12 \\
NGC 6254 &   M10 & -1.56 & 77 & 5.21 &-0.40 & 0.40 & 0.70 &-0.80 & 0.27 & 1.54 & 14.9 &5.44 \\
NGC 3201 &       & -1.50 & 94 & 5.21 &-0.30 & 0.30 & 0.60 &-0.80 & 0.34 & 1.12 & 20.2 &6.72 \\
NGC 5904 &    M5 & -1.34 &106 & 5.75 &-0.20 & 0.45 & 0.70 &-0.80 & 0.38 & 1.73 & 24.5 &7.59 \\
NGC 6218 &   M12 & -1.31 & 66 & 5.15 &-0.20 & 0.50 & 0.80 &-0.80 & 0.34 & 1.62 & 20.9 &6.87 \\
 NGC 288 &       & -1.23 & 64 & 4.92 &-0.10 & 0.25 & 0.80 &-0.80 & 0.29 & 1.35 & 16.5 &5.83 \\
NGC 6121 &    M4 & -1.20 & 80 & 5.10 &-0.10 & 0.40 & 0.80 &-0.80 & 0.33 & 0.88 & 19.3 &6.52 \\
NGC 6171 &  M107 & -1.06 & 27 & 5.07 & 0.00 & 0.40 & 0.80 &-0.80 & 0.31 & 2.01 & 17.7 &6.14 \\
NGC 2808 &       & -1.10 & 90 & 5.98 &-0.10 & 0.40 & 0.60 &-0.80 & 0.42 & 3.14 & 29.0 &8.40 \\
NGC 6838 &   M71 & -0.80 & 31 & 4.46 & 0.00 & 0.50 & 0.80 &-0.80 & 0.25 & 0.93 & 13.3 &4.98 \\
 NGC 104 & 47 Tuc & -0.74 &109 & 5.99 & 0.10 & 0.35 & 0.75 &-0.80 & 0.40 & 1.66 & 27.2 &8.10 \\
NGC 6388 &       & -0.40 & 29 & 5.99 & 0.00 & 0.20 & 0.70 &-0.80 & 0.39 & 5.56 & 26.1 &7.90 \\
NGC 6441 &       & -0.34 & 24 & 6.08 &-0.10 & 0.20 & 0.70 &-0.80 & 0.36 & 9.16 & 23.0 &7.30 \\
\enddata
\tablecomments{(1): average GC [Fe/H] abundance; (2): number of stars
  used in the analysis; (3): logarithm of GC stellar mass in units of
  $\Msun$; (4): [Na/Fe] abundance of normal material; (5): [O/Fe]
  abundance of normal material; (6): [Na/Fe] abundance of pure
  processed material; (7): [O/Fe] abundance of pure processed
  material; (8): AGB ejecta mass fraction; (9): average
  cluster-centric radius of stars in units of the cluster half-mass
  radius; (10): first generation mass enhancement factor; (11): total
  mass enhancement factor.}
\label{t:data}
\end{deluxetable*} 

Recently, Carretta and collaborators have published abundance
measurements for $>1800$ red giant stars in 20 GCs spanning a wide
range in mass and metallicity \citep{Carretta06, Carretta07a,
  Carretta09b, Carretta10b}.  The abundances were derived from high
resolution spectra obtained with the FLAMES spectrograph at the VLT.
Basic data for this sample are collected in Table \ref{t:data},
including common names, average [Fe/H] values, and number of stars
with abundance determinations or upper limits for both [O/Fe] and
[Na/Fe].  Stellar masses are derived from luminosities adopted from
the \citet{Harris96} catalog assuming $M/L_V=2 \Msun/L_\odot$.  The
table also includes a variety of derived products discussed in detail
below, including the inferred [Na/Fe] and [O/Fe] abundances of both
processed and normal material, the fraction of processed material (AGB
ejecta), the average distance of the stars observed to the cluster
center in units of the half-mass radius (the latter adopted from the
Harris catalog), and the mass enhancement factors.

Stars were used in the following analysis only if they have reported
abundance measurements or upper limits for both [Na/Fe] and [O/Fe].
Some stars were observed in a configuration that did not include the
spectral region covering the [OI] line; such stars are not included
in the analysis below.  The decision of whether or not to include the
[OI] line is not correlated with the actual [O/Fe] abundance, so
removing such objects should not affect the results.

The model [Na/Fe] and [O/Fe] abundances of the normal and processed
material for each cluster were estimated by eye from the distribution
of stars in the [Na/Fe]-[O/Fe] plane.  The estimated abundances of the
normal material lie within the range of the abundances of field stars
at the same [Fe/H] \citep{Venn04, Carretta10c}.

\begin{figure*}[!t]
\center
\resizebox{7.3in}{!}{\includegraphics{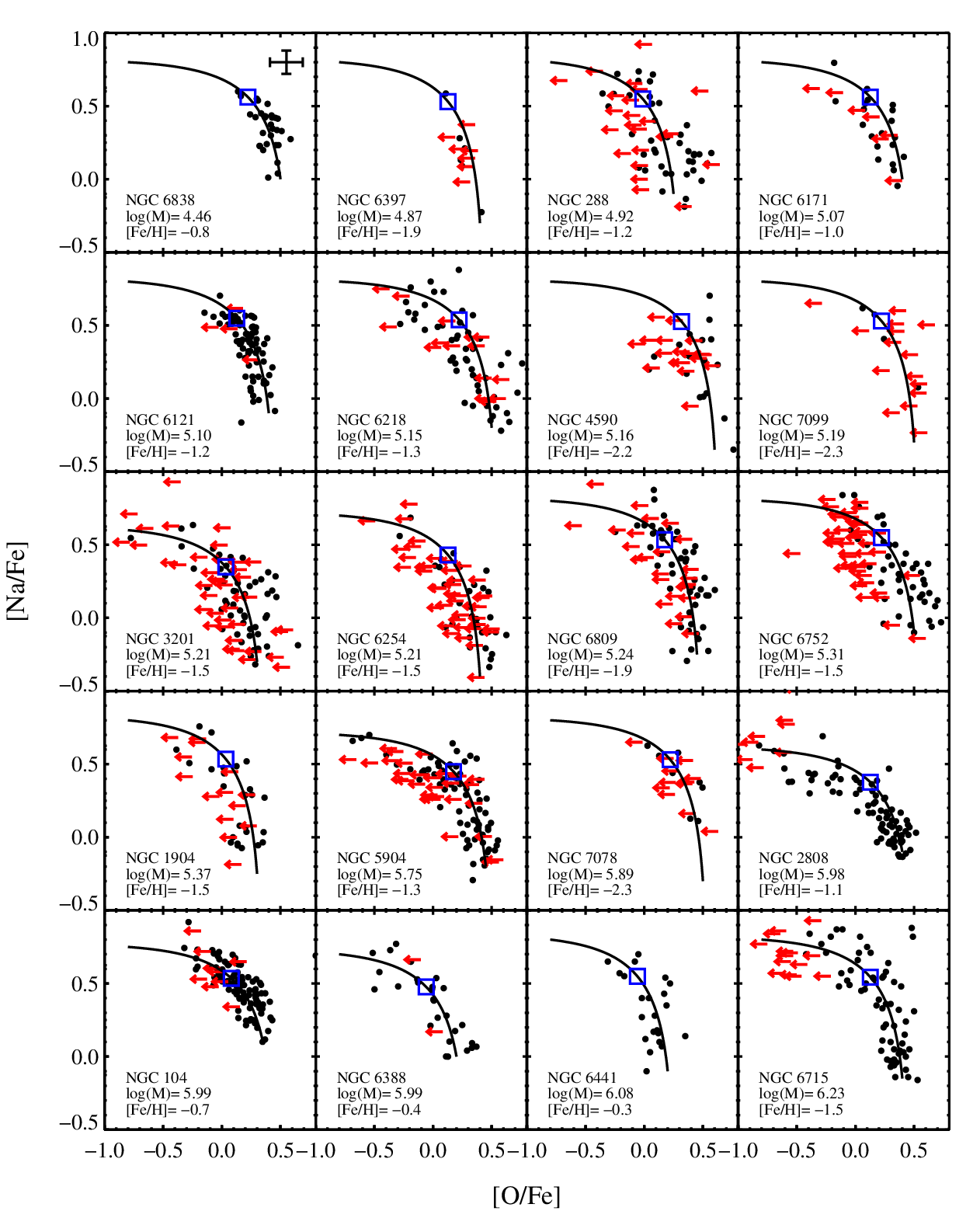}}
\vspace{0.2cm}
\caption{Na-O anti-correlation for 20 GCs.  Data are from
  \citet{Carretta06}, \citet{Gratton07}, and \citet{Carretta07a,
    Carretta07b, Carretta09b, Carretta09c, Carretta10b}.  GCs are
  sorted by mass, and the logarithm of the GC stellar mass in solar
  units is shown in each legend, as is the average [Fe/H] of each
  cluster.  Arrows indicate upper limits on [O/Fe] abundances.  A
  typical error on the abundances is shown in the upper left panel.
  Lines show the dilution models, and open squares mark the location
  at which the contribution from AGB ejecta and normal material is
  equal.}
\label{f:nao}
\end{figure*}

The distribution of [Na/Fe] and [O/Fe] abundances is shown for all 20
GCs in Figure \ref{f:nao}, along with the corresponding dilution
models.  The typical error on the [Na/Fe] and [O/Fe] abundances is
0.08 and 0.14 dex, respectively \citep{Carretta09b}.  These errors are
dominated by uncertainties in the measured equivalent widths, but also
include error due to the adopted atmospheric parameters.  The GCs are
sorted by increasing stellar mass.  It is apparent that the lower mass
GCs have on average shorter Na-O anti-correlations than the higher
mass clusters, a trend that has been noticed before
\citep[e.g.,][]{Carretta07c, Carretta10c}.  This trend, readily
apparent in the data, will be the ultimate source of the GC
mass-dependent trends discussed below.

\subsection{Derivation of $f_p$ for Galactic GCs}

\begin{figure}[!t]
\center
\resizebox{3.5in}{!}{\includegraphics{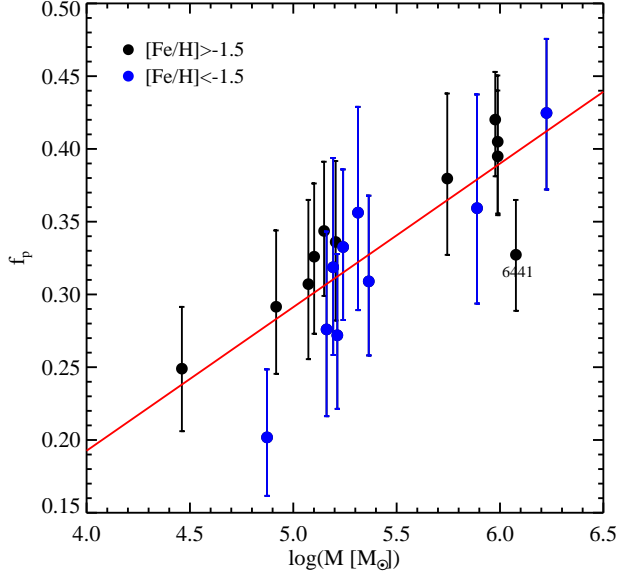}}
\caption{Relation between the fraction of GC mass comprised of AGB
  ejecta, $f_p$, and GC stellar mass.  GCs are color-coded according
  to their metallicity.  The solid line is the best-fit linear
  relation.  The metal-rich bulge GC NGC 6441 is labeled.}
\label{f:fproc}
\vspace{0.1cm}
\end{figure}

By comparing the abundance patterns of stars within each GC to the
corresponding dilution model, $f_p^j$ for each star and hence the
average $f_p$ for the cluster can be estimated\footnote{For stars with
  upper limits on [O/Fe] abundances, only the [Na/Fe] values were used
  to determine $f_p^j$.  Other choices for how to handle upper limits
  do not change the overall results.}. The result is shown as a
function of GC stellar mass in Figure \ref{f:fproc}.  The parameter
$f_p$ is only weakly sensitive to the adopted abundances of the pure
processed and normal material in the dilution model, as discussed
below.  Errors on $f_p$ have been estimated in the following way.  New
dilution models were run for each GC with variation in [O/Fe]$_p$ by
$\pm0.4$, [Na/Fe]$_p$ by $\pm0.1$, [O/Fe]$_o$ by $\pm0.05$, and
[Na/Fe]$_o$ by $\pm0.1$, and the resulting minimum and maximum $f_p$
values from this procedure were adopted as the standard error.  The
range adopted for each parameter was determined by the typical
variation allowed by the data.  The [Na/Fe]$_p$ abundance is
relatively well-constrained by the data, especially for GCs with stars
both above and below the knee in the model.  The choice of [O/Fe]$_p$
has only a modest effect on $f_p$ because of the nature of the
dilution model - stars on the upper branch of the Na-O correlation can
have very different [O/Fe] values and yet very similar $f_p^j$.

There is a strong correlation between GC mass and the fraction of mass
comprised of pure AGB ejecta.  The correlation is particularly strong
when restricted to GCs with [Fe/H]$>-1.5$.  This correlation is a
quantitative manifestation of the trend noted above that low mass GCs
have short Na-O anti-correlations compared to high mass GCs.  The
best-fit relation between $f_p$ and GC mass is:
\noindent
\be f_p = (0.10\pm0.021 )\,{\rm log}(M/M_{\odot})-(0.21\pm0.12).
\ee
\noindent
There are only two GCs at the low mass end, and one might wonder to
what extent they are influencing the observed correlation.  If those
two clusters are removed, the significance of the correlation is
reduced from $4.8\sigma$ to $3.1\sigma$; the trend is therefore robust
to the removal of the two lowest mass clusters in the sample.

\begin{figure}[!t]
\center
\resizebox{3.5in}{!}{\includegraphics{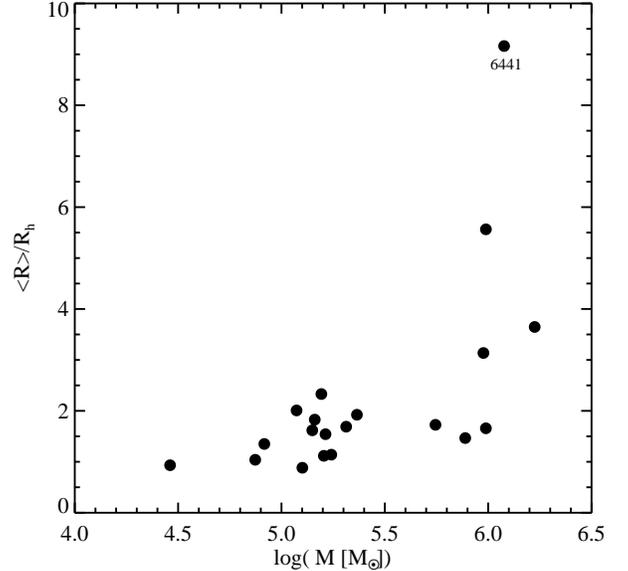}}
\caption{Average cluster-centric radius of stars for which abundance
  measurements are available, in units of the GC half-mass radius.
  Results are shown for all 20 GC as a function of GC mass.  There is
  no strong trend with GC mass, which implies that the results in
  Figure \ref{f:fproc} are not biased by observing clusters of
  different masses at different average radii.  The metal-rich bulge
  GC NGC 6441 is labeled as it is a significant outlier.}
\label{f:raverh}
\vspace{0.1cm}
\end{figure}

As discussed in the Introduction, there is some evidence that the
extent of the Na-O anti-correlation is a function of radius within
GCs.  One might then wonder if the trend seen in Figure \ref{f:fproc}
is due to a selection effect.  In order to test for such an effect,
for each GC the average radius of stars for which abundance data are
available was computed in units of the GC half-mass radius.  The
results are tabulated in Table \ref{t:data} and shown as a function of
GC mass in Figure \ref{f:raverh}.  The lowest-mass GC and the four
GCs with $\langle R\rangle/R_h>3$ define a weak trend with mass.  If
these five GCs are excluded, there is no remaining trend with mass in
Figure \ref{f:raverh}, and yet the strong correlation between $f_p$
and mass remains.  Future observations over a greater range in radius
would be valuable, but even with current data it is clear that the
trend of $f_p$ with mass is not due to sampling systematically
different regions of GCs as a function of their mass.

As can be seen in Table \ref{t:data}, the adopted pure AGB ejecta
abundances [O/Fe]$_p$ and [Na/Fe]$_p$ vary only weakly, or not at all,
with GC mass and [Fe/H].  The abundances were chosen to be
[O/Fe]$_p=-0.8$ and [Na/Fe]$_p=0.8$ initially, and only changed if
demanded by the data.  The adopted abundances for each GC are not
unique, and for the GCs with the shortest Na-O anti-correlations,
there is a substantial amount of room for variation.  

The pure processed abundances could have been adopted from the latest
AGB yields of \citet{Ventura09}, but, as noted by those authors and
others \citep[e.g.,][]{Ventura10b}, the yields are quite uncertain,
especially for Na and O, and do not provide good fits to the observed
abundances of anomalous stars.  For example, \citet{DErcole12} were
able to model the very low [O/Fe] abundances in NGC 2808 only after
invoking an additional deep-mixing mechanism in red giants to lower
the [O/Fe] abundances below the standard AGB yields.  It is clear that
AGB models still cannot, from first principles, produce elemental
yields that match even the {\it range} of observed abundances in the
second generation of stars in GCs.  It is for this reason that the
model AGB yields were not adopted when choosing [O/Fe]$_p$ and
[Na/Fe]$_p$.

The pure AGB ejecta abundance ratios could also have been chosen for
each GC to coincide with the most anomalous observed stars in each
cluster.  This would result in an increase in $f_p$ for each cluster.
Consider for example NGC 6838, which as can be seen in Figure
\ref{f:nao}, contains a short Na-O anti-correlation.  If the pure AGB
ejecta abundances were chosen to be [O/Fe]$_p=0.0$ and [Na/Fe]$=0.6$,
then the resulting pure processed fraction would be $f_p=0.4$ rather
than $f=0.25$.  Doing this for each GC would yield a pure processed
fraction, $f_p$ that was essentially independent of GC mass.

However, there is no known reason why the processed yields should be a
function of GC mass at fixed [Fe/H].  Compare for example NGC 2808
with NGC 6171.  These GCs differ by a factor of ten in mass but have
nearly the same [Fe/H].  The adopted pure processed abundances in NGC
6171 can be justified based on the fact that the same abundances fit
the whole extent of the Na-O anti-correlation in NGC 2808.  Comparing
other GC pairs with similar metallicities, such as NGC 6838 and NGC
104, NGC 6397 and NGC 6809, or NGC 6218 and NGC 5904, provides
additional justification for the adopted pure processed yields.

In summary, the derived pure AGB ejecta fractions, $f_p$, and
especially the trends with GC mass, are robust to the details of the
model, but they do depend on the assumption that the AGB yields are
independent of GC mass at fixed [Fe/H].

\subsection{Characterization of the Multiple Stellar Population
  Phenomenon in Galactic GCs}

\begin{figure*}[!t]
\center
\resizebox{6.7in}{!}{\includegraphics{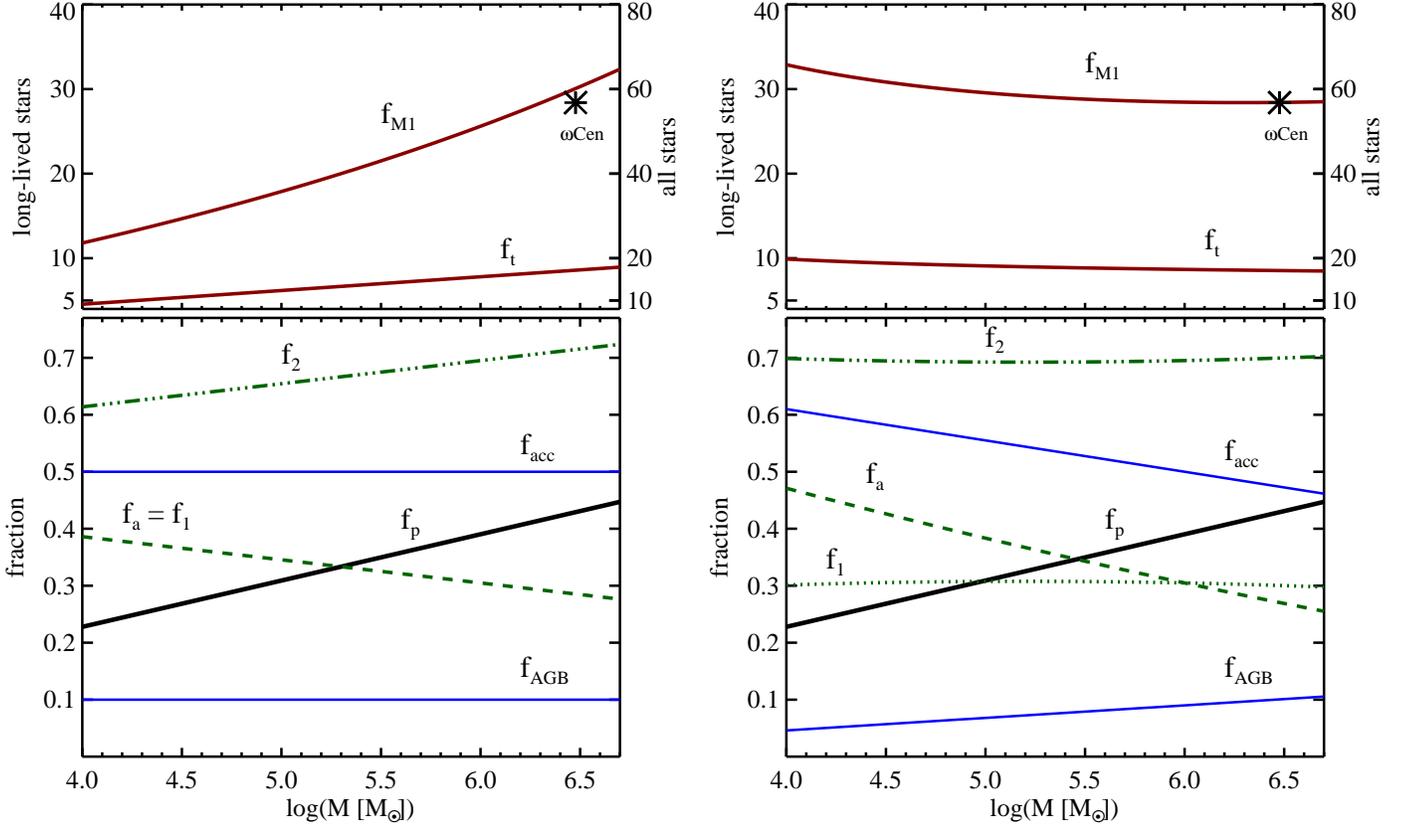}}
\caption{Characterization of the multiple stellar population
  phenomenon in GCs.  Properties shown include the fraction of present
  GC mass in pure AGB ejecta, $f_p$, in accreted material, $f_a$, in
  first generation stars, $f_1$, in the second stellar generation,
  $f_2=f_p+f_a$, fraction of second generation mass comprised of
  pristine accreted material, $\facc$, fraction of initial mass that
  ends up in AGB ejecta and is available for second generation star
  formation, $\fagb$, mass enhancement factor in the first generation,
  $f_{M1}$, and total mass enhancement factor, $f_t$.  A star
  formation efficiency of $\esf=0.5$ is assumed.  The thick black line
  is derived from the observations (see Figure \ref{f:fproc}), the
  blue lines are assumptions, and the red and green lines follow
  directly from $\esf$, $\fagb$, $\facc$, and $f_p$.  In the upper
  portion of the figures, the $y-$axis on the left side describes the
  enhancement factors of long-lived stars, while the right side
  describes the enhancement including all stars.  The approximate
  location of $f_{M1}$ for $\omega$Cen is indicated.  {\it Left
    panel}: The parameters $\fagb$ and $\facc$ are assumed to be
  constant.  {\it Right panel:} The mass-dependence of the parameters
  $\fagb$ and $\facc$ are chosen such that the majority of the other
  derived functions are independent of mass.}
\label{f:fall}
\vspace{0.1cm}
\end{figure*}

The relation between $f_p$ and GC mass derived in the previous section
can now be used to derive all of the other quantities described in
Table \ref{t:frac}, once the parameters $\esf$, $\facc$ and $\fagb$ are
specified.  Standard stellar evolution in conjunction with a
\citet{Kroupa01} IMF suggests $\fagb\approx0.1$ for AGB stars in the
$3-8\Msun$ range, assuming that all of the AGB ejecta is later
available for second generation star formation.  Variation of this
parameter and $\facc$ will be discussed below.  A star formation
efficiency of $\esf=0.5$ is also adopted.

Relations between various derived properties and present GC mass are
shown in Figure \ref{f:fall}.  Properties include the fraction of
present GC mass in pure processed material, $f_p$, in accreted
material, $f_a$, in first generation stars, $f_1$, in the second
stellar generation, $f_2$, and mass enhancement factor of stars in the
first generation, $f_{M1}$, and total mass enhancement factor, $f_t$.

The left panel shows results assuming constant values for $\fagb$ and
$\facc$.  The right panel shows how the results change when these two
parameters are allowed to be mass-dependent.  The mass-dependence of
these parameters was chosen so as to produce approximate
mass-independence of the derived functions $f_{M1}$, $f_t$, $f_2$, and
$f_1$.  The variation in results between the left and right panels
serves to illustrate the approximate range in these parameters allowed
by the data given the model uncertainties.

It is entirely plausible that both $\fagb$ and $\facc$ are in reality
mass-dependent.  For example, if Bondi accretion is the physical
process responsible for bringing in the accreted material, then
$\facc$ should increase with mass, while if simple geometric cross
sectional sweeping were dominant, then it may decrease with increasing
GC mass \citep[see e.g.,][]{Conroy11b}.  The parameter $\fagb$
represents the fraction of stellar mass that is returned to the ISM as
AGB ejecta and available for second generation star formation.  If an
increasing fraction of AGB ejecta is lost from the young cluster at
lower GC masses, then $\fagb$ could decrease with decreasing mass.
Such a scenario seems plausible since the wind velocity from AGB stars
is in the range $10-20\,\kms$ \citep{Loup93, Vassiliadis93}, which is
of order the escape velocity for low mass GCs.

The approximate value of $f_{M1}$ is shown for $\omega$Cen, and is
estimated as follows.  \citet{Renzini08} estimates that 0.7\% of the
initial mass of a stellar population would be returned to the ISM as
{\it fresh} He (i.e., He produced within stars, as opposed to
primordial He) from AGB stars with $3<M<8\Msun$.  In $\omega$Cen,
approximately 57\% of the stars belong to the metal-poor component,
33\% to the intermediate-metallicity component, and 10\% to the
metal-rich component \citep{Piotto05}.  The CMD morphology of these
components suggests He mass fractions of $Y=0.25$, $Y\approx0.38$, and
$Y\sim0.4$; the latter value is highly uncertain \citep{Sollima05}.
Taken together, this implies a total mass in fresh He of
$\approx1.7\times10^5\Msun$ in long-lived stars.  If the metal-poor
population represents the first generation, then it could have
produced only $1.2\times10^4\Msun$ of fresh He, assuming a total mass
in long-lived stars of $3\times10^6\Msun$ for $\omega$Cen.  Assuming
$\esf=0.5$, the mass enhancement factor for the first generation is
therefore $f_{M1}=28$.  This independent estimate of $f_{M1}$ agrees
fairly well with the value expected for its mass based on
extrapolation of the model presented herein.

Several generic features stand out in Figure \ref{f:fall}.  First, the
fraction of second generation stars is always $>50$\%, and is only a
weak function of mass \citep[see also][]{Carretta10c}.  While $f_2$ is
relatively constant, the composition of the second generation stars
(given by $f_p$ and $f_a$) varies more strongly with mass, with the
most massive GCs harboring second generation stars comprised largely
of pure AGB ejecta, and lower mass GCs containing second generation
stars comprised mostly of normal material accreted from the ambient
ISM.  This is a natural consequence of the fact that lower mass GCs
have shorter Na-O anti-correlations compared to higher mass GCs.

Second, the mass enhancement factor for first generation stars is very
large, approximately 30 for the most massive clusters, and at least 10
for the less massive ones.  The total mass enhancement factor, $f_t$,
is also quite large, though by factors of $2-3$ smaller than $f_{M1}$.
Previous work drew attention to the fact that $f_{M1}$ has to be large
in order to explain the observed abundance patterns, and therefore
concluded that GCs must have been substantially larger at birth
compared to their present masses.  But this is only true with regards
to the first generation population.  Since the first generation is
subdominant by number at the present epoch, the ratio of total mass at
birth to the total present mass is factors of several smaller than
$f_{M1}$.

In the present analysis attention has been focused on the mass
enhancement factors for {\it long-lived} stars.  However, short-lived
stars (defined as stars with ages less than the present age of the
Universe) comprise roughly one half of the total initial mass in a
coeval stellar population formed from a \citet{Kroupa01} IMF.  The
mass enhancement factors for {\it all} stars are therefore a factor of
approximately two larger than that quoted for long-lived stars only.
For the most massive clusters, the total mass enhancement factors for
all stars are therefore $\approx20$.

It is important to recognize that these mass enhancement factors are
all strictly {\it lower limits} because any mass-loss mechanism that
affects both first and second generation populations equally will not
affect the distribution of stars in the Na-O plane, which is the
fundamental observable underpinning the present discussion.  For
example, lower mass clusters (several $10^5\Msun$) can lose a
significant fraction of their mass via stellar evaporation over a
Hubble time \citep{Fall01}.

The basic conclusion from this section is that the ancient Galactic
GCs had to be at least $10-20$ times more massive at birth in order to
produce enough AGB ejecta to account for the observed distribution of
[Na/Fe] and [O/Fe] abundances.

Finally, the ratio $f_p/f_2$ quantifies the fraction of mass in second
generation stars comprised of pure AGB ejecta.  This ratio is
$30$\%$-60$\% over the full range in GC mass.  Broadly speaking,
second generation stars are comprised of half pure AGB ejecta and half
pristine material, in agreement with previous modeling of the
elemental abundance variations within GCs \citep{DErcole10b}.

\begin{figure}[!t]
\center
\resizebox{3.5in}{!}{\includegraphics{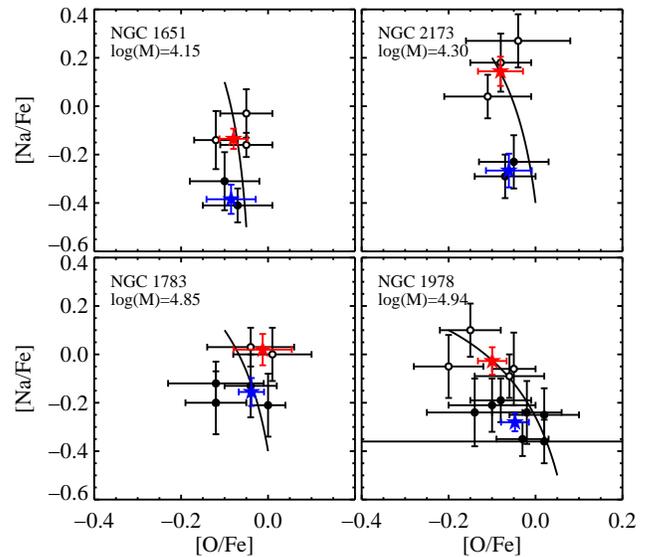}}
\vspace{0.1cm}
\caption{Na-O anti-correlations for intermediate-age LMC clusters
  sorted by cluster mass.  Data are from \citet{Mucciarelli08}.
  Stellar masses are estimated as described in the text and are shown
  in the legends in solar units.  Dilution models are shown as lines.
  For each cluster, the data are subdivided by the mean sodium
  abundance.  The mean abundance pattern for each subpopulation is
  shown as stars.}
\label{f:lmc}
\vspace{0.1cm}
\end{figure}

\section{Elemental Abundance Variations in LMC Clusters}

Knowledge of the abundance variations in star clusters outside the
Galaxy is limited because of the substantial distances to even our
nearest neighbors harboring clusters, the Magellanic Clouds.
Recently, \citet{Mucciarelli09} measured Na and O abundances for 18
giants in three old metal-poor GCs in the LMC.  They found clear
evidence for an Na-O anti-correlation in these clusters that looks
broadly similar to the Na-O anti-correlations observed in Galactic
GCs.

However, \citet{Mucciarelli08} measured elemental abundances in 27 red
giant stars located within four intermediate-age ($1-2$ Gyr),
moderate-metallicity LMC clusters, and concluded that the abundance
patterns ``show negligible star-to-star scatter within each cluster''.
Recall that these and other intermediate-age LMC clusters show
evidence in their main sequence turn-off points for an internal age
spread of several $10^8$ yr \citep{Goudfroij09, Milone09}.  These
results have been interpreted by \citet{Bekki11} as evidence for a
qualitatively different scenario at work in these clusters (wherein
these clusters capture giant molecular clouds that provide the fuel
for secondary star formation, without AGB ejecta).  It would be much
more appealing, on the principle of simplicity, if the multiple
stellar population phenomenon observed in these intermediate-age LMC
clusters were similar to what is conjectured to have occurred in the
ancient Galactic and LMC GCs.  A thorough understanding of these
clusters is also essential if one hopes to observe present-day young
clusters in the process of forming second generation stars.

Motivated by these considerations, the abundance data presented in
\citet{Mucciarelli08} are reinterpreted in the context of the dilution
models presented herein.  Figure \ref{f:lmc} shows the Na-O
anti-correlations for the four intermediate-age LMC clusters.  Each
cluster has an average [Fe/H] abundance of -0.3 to -0.5, which, for
solar abundance ratios, corresponds to
$Z=6\times10^{-3}-1\times10^{-2}$.  The stellar masses of these
clusters are estimated by adopting $V-$band magnitudes from
\citet{vandenBergh81}, an average $E(B-V)=0.08$, and a distance
modulus of $18.5$.  A $V-$band mass-to-light ratio of 0.4 was adopted,
which is appropriate for a stellar age of 1.6 Gyr and
$Z=6\times10^{-3}$.

The stars within each cluster have been split into two groups
according to the mean sodium abundance of the stars.  The mean sodium
and oxygen abundances were then computed within each subpopulation.
In all clusters there is statistically-significant evidence for
different mean sodium abundances in the two populations.  The two
populations are most visually striking in NGC 2173 and NGC 1978.  Of
course, computing the mean abundance of subpopulations split according
to that abundance will tend to exaggerate the differences between the
two populations.  Another way of assessing the statistical
significance of the internal sodium variation is to consider the stars
at the extremes of the distribution.  In NGC 1783 the stars with the
highest and lowest [Na/Fe] abundances are consistent within $1\sigma$,
while for NGC 1651, NGC 2173, and NGC 1978 the most extreme stars
differ at the $\gtrsim3\sigma$ level.  There is therefore strong
evidence for internal variation in [Na/Fe] in the clusters NGC 1651,
NGC 2173, and NGC 1978, and ambiguous or weak evidence for NGC 1783.

The [O/Fe] abundances show considerably less variation, and it is
perhaps for this reason that \citet{Mucciarelli08} concluded that
there was negligible star-to-star variation within each cluster.
However, two important facts conspire to reduce the expected variation
in [O/Fe] in clusters that contain second generation star formation
from AGB ejecta.  First, the first generation stars in the
intermediate-age clusters are not $\alpha-$enhanced, and so the
maximum value of [O/Fe] is reduced to roughly 0.0.  Second, at the
moderate metallicities characteristic of these clusters, the depletion
of O abundances due to hot bottom burning in AGB stars is much less
than at lower metallicities characteristic of the ancient clusters.
The mass- and metallicity-dependent AGB yields from the models of
\citet{Ventura09} make this point clearly.  In their
$Z=4\times10^{-3}$ models, the [O/Fe] yields for AGB stars with masses
in the range $3-6.5\Msun$ never drop below 0.0.  So while the full
range in [O/Fe] for the low metallicity GCs routinely exceeds 1 dex,
at higher metallicities the range is expected to collapse to nearly
zero, as observed.

In Figure \ref{f:lmc}, schematic dilution models are included to guide
the eye.  Only NGC 1978 shows signs of a true Na-O anti-correlation.
It is intriguing that this is also the most massive intermediate-age
cluster in the sample.  However, this is the only massive cluster
studied by \citet{Milone09} where a significant spread in its main
sequence turn-off point was {\it not} detected, suggesting a coeval
population.  As noted in that work, the CMD data for this cluster were
obtained in the cluster outskirts, unlike the other clusters studied.
The lack of an observed age spread could therefore be explained if the
second generation is more centrally concentrated than the first
generation, as models predict.

While it is clear from the data that elemental abundance variations
exist in the intermediate-age LMC clusters, at least for [Na/Fe], it
should not be surprising if the range of the variation is smaller than
in the Galactic GCs (even setting aside the previous discussion of
[O/Fe]).  The trend of $f_p$ with mass found for Galactic GCs suggests
that lower mass GCs are able to retain only a fraction of the AGB
ejecta from first generation stars.  If these LMC clusters were not
dramatically more massive in their past, then the expected range in
the elemental abundance variations would be smaller, as the shallower
potential wells would not be able to retain as much AGB ejecta.  The
point is that one should not expect clusters in other environments to
show the exact same abundance variations as observed in Galactic GCs.
However, it {\it is} expected that any cluster showing an internal
variation in age, as seen for these LMC clusters, should also show
{\it some} variation in their light element abundances, as observed.

Two important conclusions emerge from this section.  First,
statistically-significant star-to-star scatter in the [Na/Fe]
abundances exists within three of the four intermediate-age LMC
clusters studied by \citet{Mucciarelli08}.  This contradicts the
conclusions of Mucciarelli et al., though they present no quantitative
argument for a lack of star-to-star scatter.  Second, a strong Na-O
anti-correlation is not expected on theoretical grounds for
moderate-metallicity clusters because the expected range in [O/Fe] is
small.  For moderate-metallicity clusters, interpretation of any
internal abundance variations must take this point into account, and
it may therefore be more profitable to focus on obtaining [Na/Fe]
abundances in such clusters.

\section{Discussion}
\label{s:disc}

\subsection{Origins \& Implications}
\label{s:origin}

The preceding analysis has revealed several important facts related to
the multiple stellar population phenomenon in GCs.  Perhaps most
important is the result that the fraction of present day GC mass
comprised of pure AGB ejecta, $f_p$, is large and strongly correlated
with GC mass.  The small scatter in $f_p$ at fixed mass suggests that
whatever process shapes the elemental abundance trends is driven
primarily by internal processes, rather than external ones such as
tides from the host galaxy.  It is puzzling why $f_p$ is so strongly
correlated with mass and yet such a weak function of mass (varying by
less than a factor of two over two decades in GC mass).

There are at least two possible explanations for the observed
correlation between $f_p$ and GC mass.  First, the amount of AGB
material retained within a GC may decrease with decreasing GC mass.
This is plausible, as lower mass GCs have lower escape velocities, and
yet the typical velocity of AGB ejecta is independent of GC mass.  A
second possibility is that GCs retain all of their AGB ejecta but this
material is more strongly diluted by pristine material accreted from
the ambient ISM in lower mass GCs.  Lower mass GCs could more
efficiently accrete pristine material if ISM sweeping is the dominant
mechanism bringing in new material \citep{Conroy11b}.  The former
explanation seems more natural and is therefore preferred.

The derived fraction of mass in second generation stars presents one
of the most significant outstanding puzzles.  It is observed to be
$>50$\% of the total present GC mass and is relatively independent of
mass.  This fraction is derived in the present work, but has been
found also by many authors through different techniques
\citep[e.g.,][]{Smith87, Carretta10c}.  If GCs were $>10$ times larger
at birth, then the early contribution of second generation stars to
the total was small, of order a few percent.  The fact that present
day GCs always end up with roughly the same fraction of second
generation stars strongly suggests that there is some mechanism that
drives all clusters toward this state.  One scenario that could
achieve this is as follows.  Imagine that some mechanism causes the
first stellar generation to become unbound on the same timescale as
the formation of the second generation (see below for an example).
The second generation, being more tightly bound, would remain so, and
might also be capable of capturing of order its own mass in first
generation stars.

The second major outstanding puzzle is the more fundamental issue of
how and why the ancient GCs were initially so much more massive at
birth.  Standard dynamical effects cannot explain this, essentially
because the relaxation time increases with increasing mass, and yet
the mass enhancement factors are larger at higher masses.  The only
somewhat plausible scenario that has been proposed to explain this is
{\it primordial} mass segregation of the stars in the first stellar
generation \citep[e.g.,][]{DErcole08}.  If a significant fraction of
high mass stars were born near the cluster center, then as they evolve
and die they will carry away a significant fraction of the total
binding energy of the young GC, causing it to expand significantly.
There is some circumstantial evidence for primordial mass segregation
in nearby young clusters whose ages are comparable to their crossing
times \citep[e.g.,][]{Hillenbrand98, deGrijs02, Stolte06, Gennaro11}.

If primordial mass segregation were strong enough, it could unbind
young GCs on a timescale of $\lesssim1$ Gyr \citep{Vesperini09}, or at
least lead to a substantial amount mass loss and cluster expansion
\citep{Marks10}.  If this were a common feature of GC evolution, then
perhaps GCs owe their very survival to the formation of tightly bound
second generation stars \citep{DAntona08b}.  These second generation
stars, born at the bottom of a deep potential well set by the first
generation stars, should be relatively immune to the effects of
primordial mass segregation.  Detailed simulations will be required to
validate this scenario, of which the simulations by \citep{DErcole08}
are an important first step.  If confirmed, this would constitute a
major revision to our understanding of the survival of massive star
clusters.

A novel implication of the fact that the ancient GCs were much more
massive at birth is their potential contribution to the Galactic
stellar halo \citep[see also][]{Vesperini10, Schaerer11}.  There are
currently $\approx150$ known GCs \citep{Harris96}, and they have an
average present mass of $10^{5.5}\Msun$.  If these GCs had on average
10 times more long-lived stars at birth, then in total they would have
equaled a mass of $\approx5\times10^8\Msun$.  This value is within a
factor of a few of the estimated stellar mass of the Galactic stellar
halo \citep[e.g.,][]{Siegel02}, and of the stellar halo in M31
\citep[e.g.,][]{Kalirai06}.  This is the contribution to the stellar
halo of GCs that remain bound at the present epoch.  \citet{Martell10}
have shown that approximately half of the Milky Way (MW) stellar halo
could have formed from GCs that are now disrupted.  An understanding
of the early evolution of GCs may therefore be required in order to
understand the hierarchical formation of the stellar halos around
galaxies.

If the average GC today was 20 times more massive at birth, then it
would have had a mass of $\approx6\times10^6\Msun$, and it would have
formed out of a giant molecular cloud (GMC) with a mass of at least
$\sim10^7\Msun$, assuming $\esf=0.5$.  Such large GMCs should form in
the massive, gas-rich protogalactic disks common at high redshift
\citep{Escala08}.  Indeed, observations of disk-dominated galaxies at
$z\sim2$ have identified large numbers of super-star forming clumps
with masses of order $10^9\Msun$ \citep{Genzel08}.  Each of these
clumps could easily spawn several massive young GCs.  Even at the
present epoch young clusters have been found with dynamical masses in
excess of $10^7\Msun$ \citep[e.g.,][]{Maraston04, Bastian06}.  While
rare at the present epoch, the conditions at high redshift may well
have favored the formation of many such massive objects.

M31 contains a large number of extended, low surface brightness GCs
with half-light radii $>10$ pc \citep{Huxor05} that have no
counterparts in the Galaxy.  The origin of these extended GCs is not
known.  In the context of the present discussion, they could arise
from the rapid expansion caused by primordial mass segregation.  If
for some reason these clusters formed in a much more benign tidal
field than the GCs in the Galaxy, then the loosely bound stars may
not have been stripped by the present epoch.  If this scenario is
correct, then these extended GCs are comprised principally of first
generation stars, and should contain a small core of second
generation stars at their center.  Radial gradients of the CN
absorption feature in these GCs would be very interesting.

The conclusion that GCs were much more massive in the past may also
require some revision to dynamical models for the long-term evolution
of the GC population \citep{Gnedin97, Fall01, Marks10}.

While the present analysis provides a self-consistent explanation for
the variation in the extent of the Na-O anti-correlation from cluster
to cluster, complications arise when considering other light elements.
In particular, some clusters that show an Na-O anti-correlation do not
harbor a corresponding Mg-Al anti-correlation \citep{Carretta09c}.
For example, the clusters NGC 6121 and NGC 6752 have very similar
[Fe/H], total mass, and $f_p$ parameters, and yet the former shows no
star-to-star variation in either Mg or Al, while the latter shows
clear variations in both.  The Mg and Al yields depend on AGB mass in
a manner different from the Na and O yields \citep{Ventura08}, and so
it is possible that differences in Mg-Al anti-correlations in GCs that
contain the same Na-O anti-correlation arise from a difference in the
average AGB mass that donates ejecta for later star formation.  See
\citet{Carretta09c} for further discussion.

\subsection{Alternative Explanations}

A number of alternative explanations have been proposed for various
aspects of the standard model considered herein.  Several of these
will be briefly discussed in this section; the reader is referred to
\citet{Renzini08} and \citet{Conroy11b} for further discussion.

While AGB ejecta is the favored source for the processed material,
other proposed sources include winds from massive ($\gtrsim 20\Msun$)
rotating stars \citep{Decressin07} and massive binary star
interactions \citep{deMink09}.  The most serious objection to these
scenarios is that they are associated with massive stars with short
lifetimes.  It is therefore not natural for the processed material to
remain free of supernovae contamination.  For example, it is not
obvious why the second stellar generation should have the same [Ca/Fe]
abundances as the first generation in these scenarios, and yet this is
observed to be so \citep{Carretta10a}.

The similar $10^8$ yr timescales for both intermediate-mass AGB stars
and the drop in UV photons from the first generation provide a simple
framework to understand not only the ancient GCs but also the observed
age spread within intermediate-age LMC clusters.  Invoking processed
material from massive stars requires either a more nuanced or an
altogether different mechanism to be at work in the LMC clusters
compared to ancient GCs.  This seems unappealing on the principle of
simplicity.

Several authors have considered the possibility that stars with
anomalous abundance ratios simply had their surfaces polluted by AGB
ejecta \citep[e.g.,][]{DAntona83, Thoul02}.  In this scenario there is
only one generation of star formation and the observed range in light
element abundances is due to the differing amount of surface pollution
from star to star.  A major advantage of this scenario is that the GCs
need not have been substantially more massive at birth because a much
smaller amount of AGB ejecta is needed to cover the surfaces of stars.

This scenario is almost certainly ruled out by the lack of strong
[O/Fe] and [Na/Fe] abundance variations between the main sequence and
red giant branch (RGB).  A $1\Msun$ star on the RGB has a convective
envelope comprising approximately 40\% of its mass, while the same
star on the main sequence has a negligible convective envelope.
Therefore, this model would predict that as such a star evolves from
the main sequence onto the RGB, surface pollution would be heavily
diluted as the convective envelope deepens.  This is not observed,
which either means that the scenario is to be discounted \citep[as
suggested in][]{Cohen02}, or the amount of surface pollution was
substantial, so that it could not be diluted even by deep convection.
The latter possibility then requires large amount of AGB ejecta, which
means that large mass enhancement factors are required.

Recently, \citet{Conroy11b} proposed a scenario for the formation of
multiple stellar populations within GCs that contains many of the
ingredients listed in the Introduction, e.g., AGB ejecta and matter
accreted from the ambient ISM as the source of second generation
stars.  The principle difference between that model and the
\citet{DErcole08} model is that the former explicitly attempted to
avoid the conclusion that the ancient GCs were much more massive at
birth.  Instead, the large amount of material needed to form the
second generation stars came primarily from the ambient ISM.  As found
in previous work and confirmed herein \citep[see also][]{DErcole11},
the difficulty with such a scenario in explaining the properties of
the ancient MW GCs is that too few anomalous stars are produced.  This
is because the accreted ambient ISM, which has normal abundance
patterns, dominates the mass budget.  This does not preclude the
possibility that star clusters in other systems formed a second
generation of stars from primarily accreted ambient ISM material.

Finally, several related scenarios appeal to highly unusual stellar
configurations in order to explain the high number of anomalous stars
\citep[e.g.,][]{Bekki06b, Bekki07, Marcolini09}.  These scenarios have
two important properties in common.  First, due to the special
configurations required, they cannot be expected to operate at the
present epoch, and therefore the multiple stellar population
phenomenon observed in intermediate-age LMC clusters requires a
second, distinct explanation.  Second, they do not require present day
GCs to have been substantially more massive at birth.

The scenario outlined by \citet{Bekki06b}, for example, envisions GC
formation at the centers of their own dark matter halos.  Stars
throughout the dark halo are allowed to donate their AGB ejecta to the
central regions where the GC is expected to form.  Unfortunately, a
number of indirect arguments disfavor GC formation at the centers of
their own dark halos \citep[see, e.g.,][for a review]{Conroy11b}, and
direct dynamical searches for dark halos around present day isolated
GCs strongly disfavor the presence of dark halos \citep{Baumgardt09,
  Conroy11a}.  Since the deep potential well provided by a dark halo
should efficiently retain supernovae ejecta, it is also not clear why
GCs in this scenario should not all have large internal variation in
heavier elements such as Fe and Ca.

\section{Summary}
\label{s:sum}

A simple model has been presented to interpret the observed Na-O
anti-correlation in 20 Galactic GCs.  The model assumes that GCs are
composed of two generations of stars: a first generation formed from
material with abundance ratios characteristic of field stars, and a
second generation formed from a mix of AGB ejecta from the first
generation and material accreted from the ambient ISM.  Principal
results from this analysis include the following:

\begin{itemize}

\item The fraction of present GC stellar mass comprised of AGB ejecta
  is strongly correlated with GC mass, varying from 0.2 to 0.45 over a
  factor of 100 in GC mass.  This result is grounded in the
  observation that the extent of the observed Na-O anti-correlation is
  strongly correlated with GC mass.

\item The fraction of GC mass in pure AGB ejecta, in conjunction with
  several well-motivated assumptions, provides strong constraints on
  the composition and mass-loss history of Galactic GCs.  The fraction
  of mass contained in second generation stars is always greater than
  50\%, for GCs ranging in mass between $10^{4.5}\Msun$ and
  $10^{6.5}\Msun$.  The population of first generation stars in GCs
  must have been factors of $20-60$ larger at birth compared to the
  present epoch.  However, owing to the fact that first generation
  stars are subdominant at the present epoch, the total GC mass was
  only factors of $10-20$ larger at birth.  These factors are
  mass-dependent and lower limits.  The ancient GCs were therefore
  much more massive at birth.

\item Elemental abundance data on four intermediate-age ($1-2$ Gyr)
  LMC clusters are reinterpreted in the context of the models
  presented herein.  It is found that three of the four clusters show
  unambiguous evidence for internal variation in [Na/Fe] abundances.
  [O/Fe] values do not show signs of star-to-star variation, but this
  is shown to be a natural expectation in moderate metallicity
  clusters.  The scenario invoked to explain the properties of the
  ancient Galactic GCs therefore appears to also be at work at the
  present epoch in the LMC.

\end{itemize}


\acknowledgments 

This work made extensive use of the NASA Astrophysics Data System and
of the {\tt astro-ph} preprint archive at {\tt arXiv.org}.  Eugenio
Carretta and Raffaele Gratton are thanked for comments on an earlier
draft.  I thank the referees for comments that have improved the
quality and clarity of the manuscript.


\end{document}